\documentclass[prb,twocolumn,amsmath,amssymb]{revtex4}
\pdfoutput=1
\usepackage{graphicx}% Include figure files
\usepackage{dcolumn}% Align table columns on decimal point
\usepackage{bm}% bold math
\newcommand{\vq}{\vec{q}}
\newcommand{\half}{\frac{1}{2}}

\begin{document}

\title{Anharmonic softening of Raman active
phonons in Iron-Pnictides; estimating the Fe isotope effect due to
anharmonic expansion.}
\author{M. Granath}
\email{mats.granath@physics.gu.se}
\affiliation{Department of Physics, University of Gothenburg,
G\"oteborg 41296 Sweden}
\author{J. Bielecki}
\author{J. Holmlund}
\author{L. B\"orjesson}
\affiliation{Department of Applied Physics,
Chalmers University of Technology,
G\"oteborg 41296
Sweden}

%\author{N. L. Wang}
%\affiliation{Beijing National Laboratory for Condensed Matter Physics, Institute of Physics,
%Chinese Academy of Sciences, Beijing 100080, P. R. China }
\date{\today}% It is always \today, today,

\begin{abstract}
We present Raman measurements on the iron-pnictide superconductors CeFeAsO$_{1-x}$F$_{x}$ and NdFeAsO$_{1-x}$F$_{x}$.
Modeling the Fe-As plane in terms of harmonic and a cubic anharmonic Fe-As interaction we
calculate the temperature dependence of the energy and lifetime of the Raman active Fe $B_{1g}$ mode and fit to the observed energy shift.
The shifts and lifetimes are in good agreement with those measured also in other Raman studies which demonstrate that the phonon spectrum, at least at
small wave numbers, is
well represented by phonon-phonon interactions without any significant electronic contribution. Even at zero temperature there is a non negligent
effect of interactions on the phonon energy, which for the Fe $B_{1g}$ mode corresponds to $6cm^{-1}$, or 3\% of the total energy of the mode.
We also estimate the anharmonic expansion from  Fe ($56\rightarrow 54$) isotope
substitution to $\Delta a \approx 5.1\cdot 10^{-4}{\AA}$ and $\Delta d_{Fe-As}\approx 2.5\cdot 10^{-4}{\AA}$ and the shift of harmonic zero point fluctuations of
bond lengths $<\Delta x^2>\lesssim 3\cdot 10^{-5}\AA^2$, giving a total relative average decrease of electronic hopping
integrals of $|\delta t|/t\lesssim 2.0\cdot 10^{-4}$. For a non-phonon mediated weak coupling superconductor this gives an isotope exponent
$\alpha\sim 10^{-2}$.
The results poses a serious challenge for any theory of superconductivity in the pnictides that does
not include electron-phonon interactions to produce a sizable Fe-isotope effect.

\end{abstract}

\maketitle

\section{Introduction}
The recently discovered iron-pnictide high temperature superconductors\cite{Kamihara} have several features in common with the much studied cuprate
superconductors. In addition to their overall quasi-two
dimensional nature such features include commensurate antiferromagnetism in close proximity to or even coexisting with superconductivity.
Intriguingly, there is also a common tendency for breaking of the fourfold rotational symmetry of the crystal\cite{Zhao} which in the case of the cuprates is most
likely an intrinsic
electronic property related to the strong correlations.\cite{Kivelson}
For the pnictides (as well as in some cases for the cuprates) there is at low doping a
crystal symmetry breaking from tetragonal to orthorombic which is closely connected to an accompanying spin density wave
(SDW) order.\cite{Dong,Cruz}
%For the cuprates it is clear that the electronic order ushers in the atomic order\cite{what} but for the pnictides this is still an
%open question.
This interplay between atomic and electronic degrees of freedom naturally leads to the question: What is the role of phonons for the electronic properties of these systems?
Focusing on the pnictides, results from density functional theory (DFT)\cite{Singh,Boeri} are credible as they
give an
electronic spectral distribution consistent with photoemission experiments\cite{ARPES} but have also found that the electron-phonon coupling is much too weak to explain the
high T$_c$.\cite{Boeri} Challenging these findings was the recent report of
an isotope shift of T$_c$ (and T$_{SDW}$) when substituting $^{56}$Fe with $^{54}$Fe with an exponent of $\alpha=-\frac{d\ln T_c}{d\ln m}=0.4$
close to the BCS value of $0.5$ for a pure
iron mode.\cite{Liu}
The mechanism for superconductivity in the pnictides is a major unresolved issue and it is clearly essential to understand better the lattice-electron interplay.

In this paper we measure and model one particular Raman active phonon which lives primarily on Fe atoms and thus common to all iron-pnictide superconductors.
The purpose is to study the anharmonic structure of the Fe-As plane with two main objectives: First, to investigate whether any non-phonon
contributions are necessary to describe the temperature dependence of
the phonon energy and lifetime. Second, to estimate the magnitude of the lattice expansion that follows from substitution with a lighter mass and the corresponding changes
in electronic hopping integrals, thus investigating the plausibility of a non-phonon related isotope effect in the pnictides.\cite{Fisher,Chakravarty}

In brief, we model the harmonic spectrum of the As-Fe plane and use standard Greens function methods to calculate the self energy of the Fe B$_{1g}$ Raman phonon
to second order in the cubic anharmonic Fe-As interatomic coupling. The harmonic spectrum is based on a minimal parameter fit to spectra derived within the
local density approximation (LDA) of DFT but the
calculation of the temperature dependent contribution from phonon-phonon interactions is beyond the capabilities of that method.
We present Raman results
on (Nd,Ce)FeAsO$_{1-x}$F$_x$ and estimate the anharmonic coupling strength by fitting our model calculations to the measured temperature dependent energy shift.
Also the width agrees well with Raman measurements on CaFe$_2$As$_2$\cite{Choi}
thus effectively ruling out any significant electronic
contribution to the broadening. Based on the magnitude of the anharmonic coupling we estimate the isotope ($^{56}$Fe$\rightarrow$$^{54}$Fe)
shift of the lattice parameters
to $\lesssim 2\cdot 10^{-4}$ and calculate a similar relative decrease of interatomic hopping integrals. Harmonic zero point fluctuations give an even smaller
isotope shift of hopping integrals. In weak coupling theory correspondingly small changes in the electronic density of states (DOS)
is expected to give an isotope exponent $\alpha\sim 10^{-2}$.

\begin{figure}
\includegraphics[width=8cm]{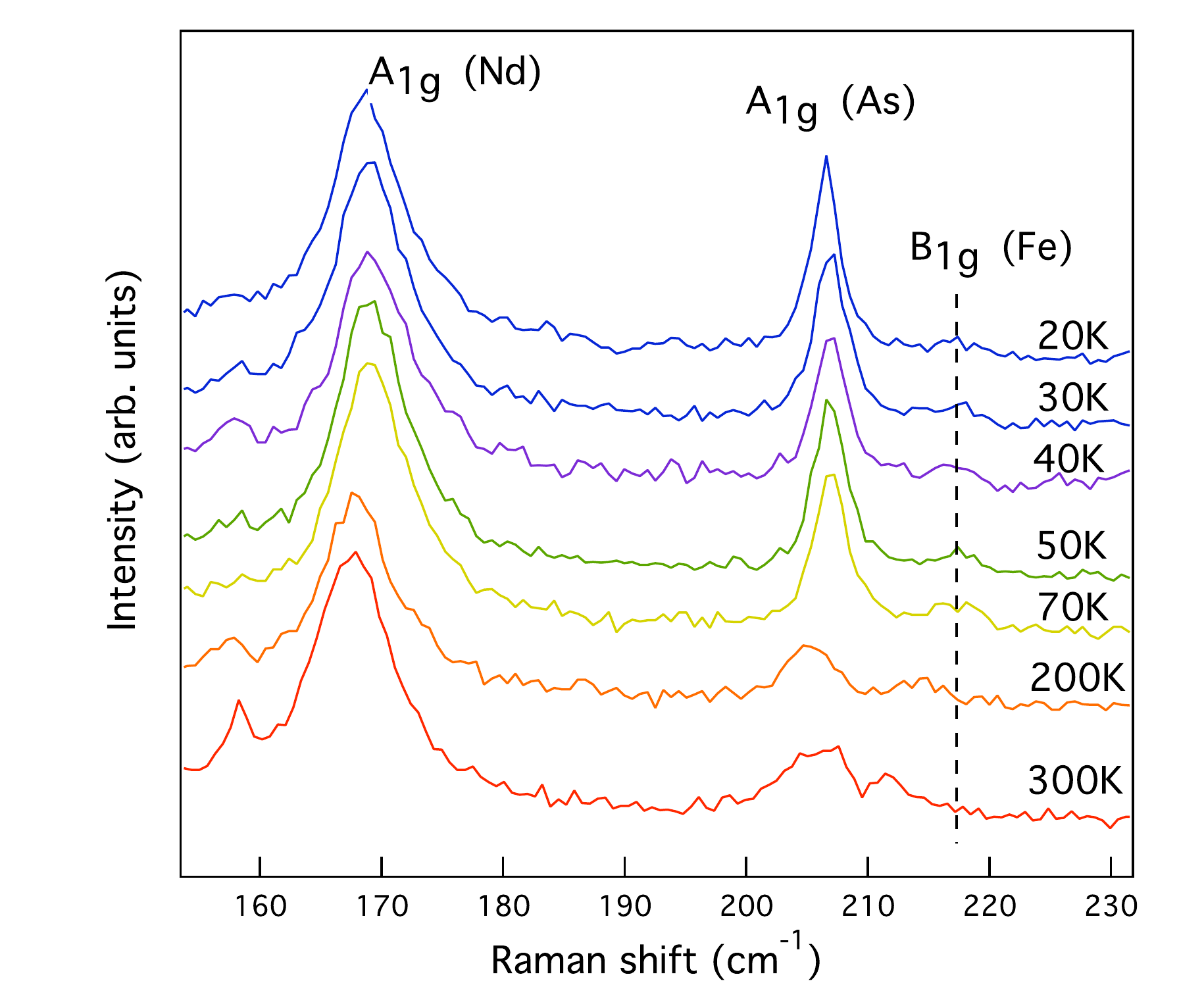}
\caption{\label{spectra_Nd} (Color online) Temperature dependent Raman spectra for NdFeAsO$_{1-x}$F$_{x}$ ($x=.12$) showing three 
Raman active phonon modes labeled by their symmetry and main atomic displacement.}
\end{figure}

\begin{figure}[b]
\includegraphics[width=8cm]{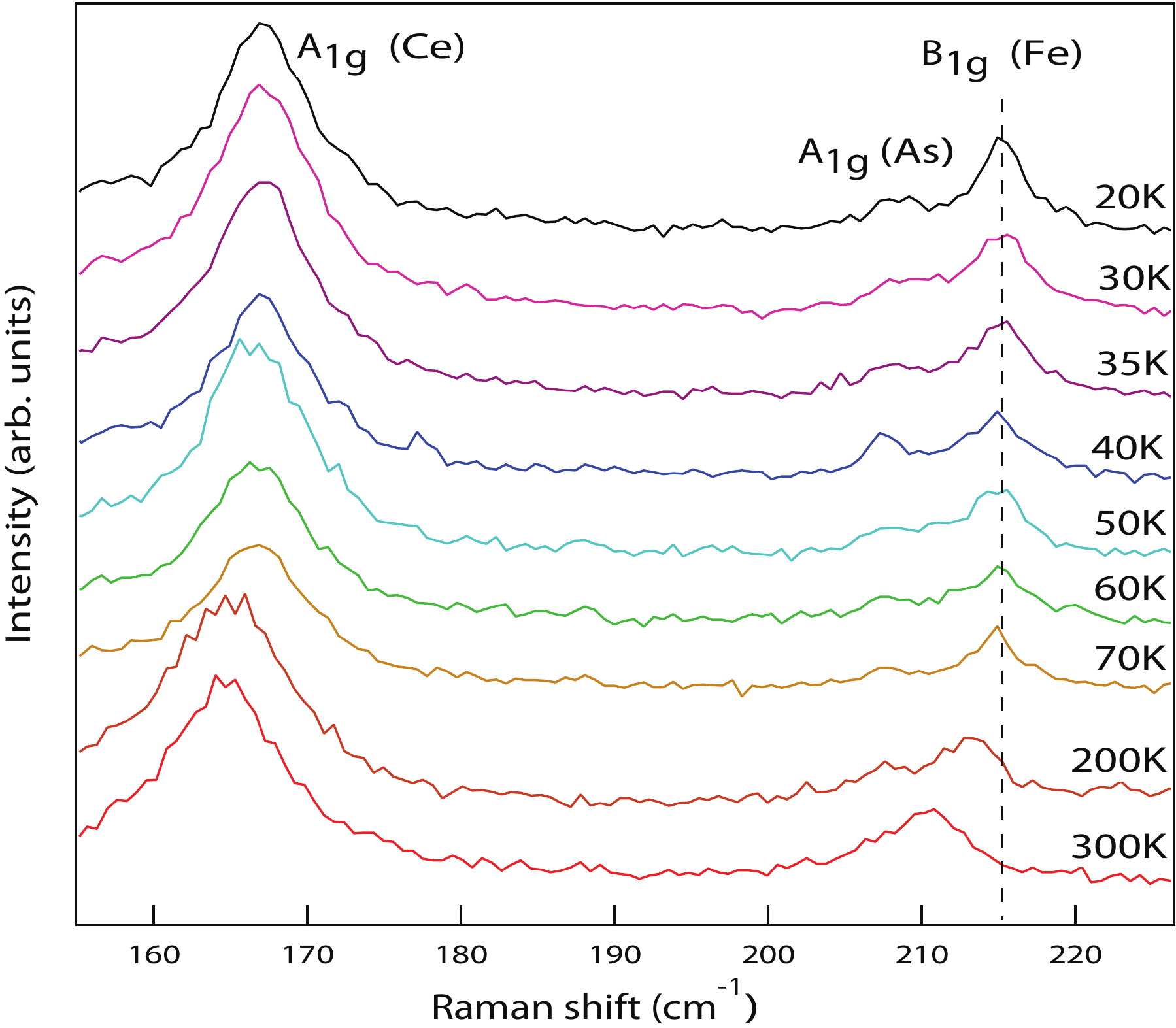}
\caption{\label{spectra_Ce} (Color online)  Raman spectra for CeFeAsO$_{1-x}$F$_{x}$ ($x=.16$).}
\end{figure}
\section{Raman spectra}

\begin{figure}[ht]
\includegraphics[width=8cm]{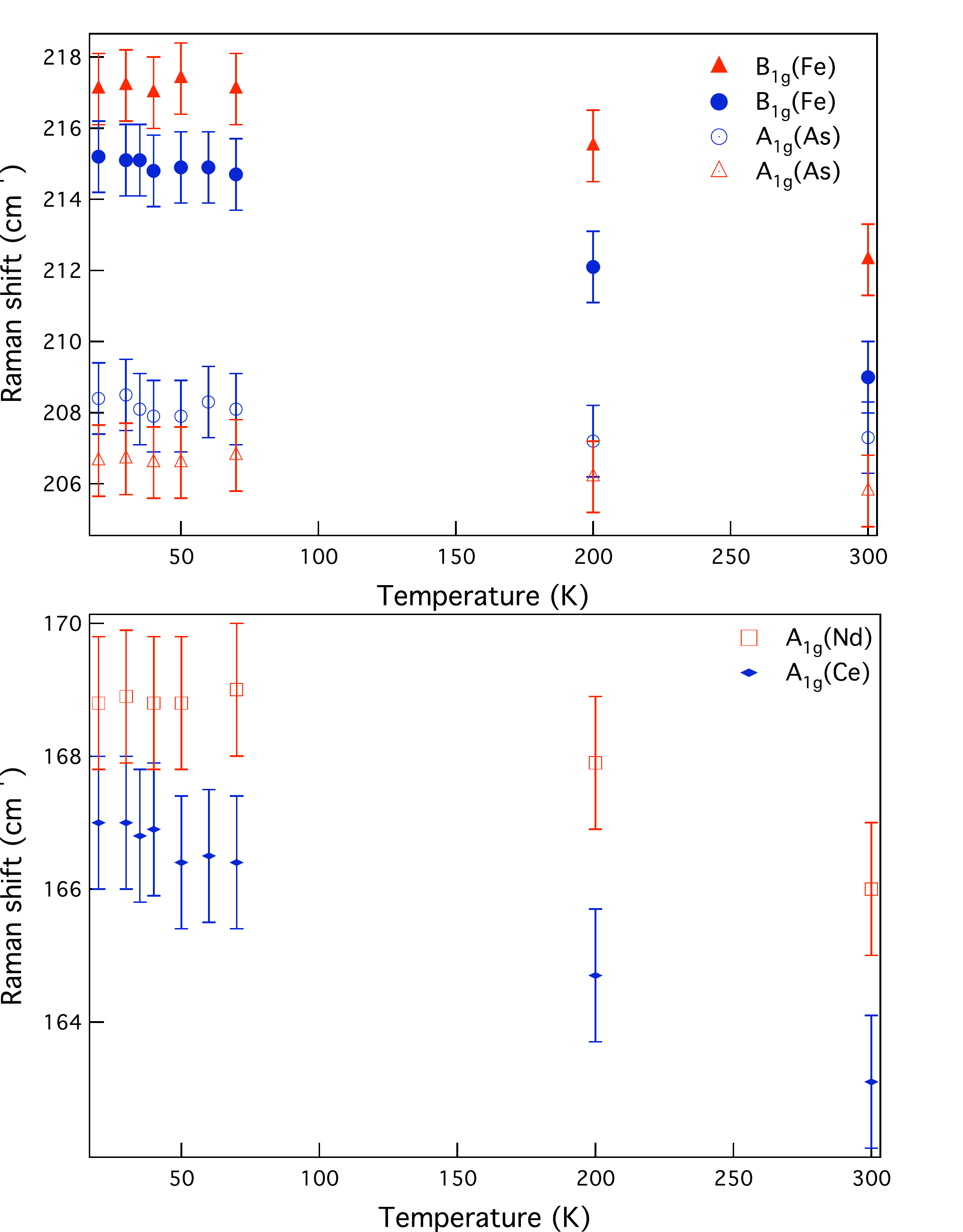}
\caption{\label{spectra_shifts} (Color online) Temperature dependence of the phonon peak position for 
NdFeAsO$_{0.88}$F$_{0.12}$ (red) as extracted from Fig. \ref{spectra_Nd} and correspondingly for 
CeFeAsO$_{0.84}$F$_{0.16}$ (blue) from Fig. \ref{spectra_Ce}.}
\end{figure}

Raman spectra between $100\rm{cm}^{-1}$ and $400 \rm{cm}^{-1}$ were collected at temperatures ranging from $20$K to $300$K for polycrystalline samples of
CeFeAsO$_{1-x}$F$_{x}$ ($x=0.16$) and NdFeAsO$_{1-x}$F$_{x}$ ($x=0.12$).\cite{note_undoped} For sample preparation and 
characterization see Chen at al.\cite{samples}
All spectra were recorded using a Dilor-XY800 spectrometer in double subtractive mode. 
In all scans the 514.5 nm line from a Ar+ laser used with a power of less than 1mW  was focused onto the samples with a spot size less than 2 $\mu$m. The samples were installed in a LHe cooled cryostat.
We observe three Raman active modes with energies around 170, 210 and 220 cm$^{-1}$, see Fig. \ref{spectra_Nd} and Fig. 
\ref{spectra_Ce}, in agreement with other
Raman studies,\cite{Raman}
that are identified as (Ce/Nd)-A$_{1g}$, As-A$_{1g}$, and Fe-B$_{1g}$ modes respectively. Also in agreement with earlier studies
we found no effect on the phonon energies from crossing into the superconducting phase (T$_c=$35K/45K for Ce/Nd).
As discussed in the introduction we will be interested in analysing the temperature dependence of the energy as 
extracted in Fig. \ref{spectra_shifts} and for modeling purposes we consider only the Fe-B$_{1g}$ ($x^2-y^2$) mode which lives primarily
in the Fe-As plane.

\section{Modeling}
%\subsection{phonon spectrum}

\begin{figure}
\includegraphics[width=8cm]{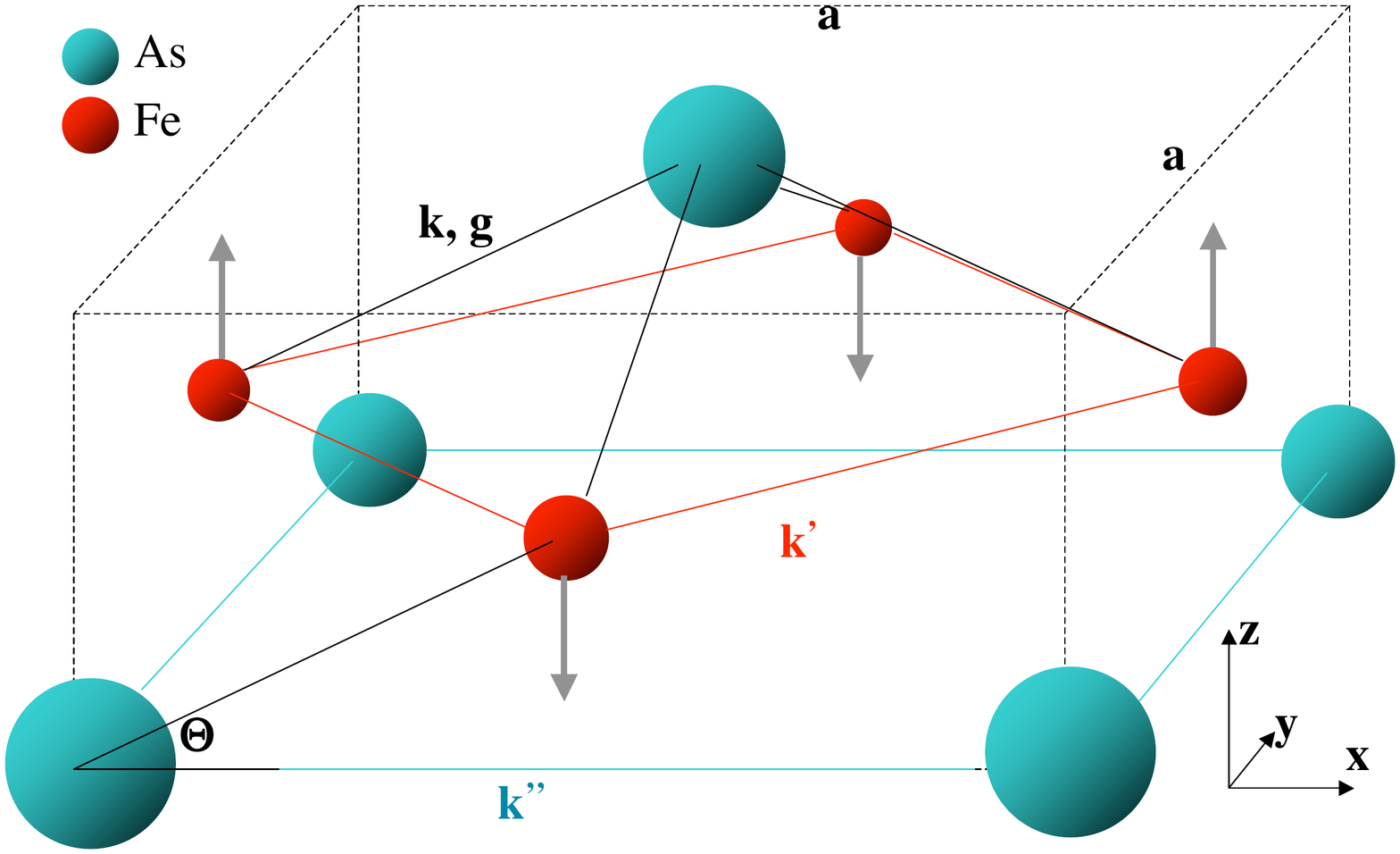}
\caption{\label{crystal} (Color online) Schematic of the Fe-As plane with in-plane lattice parameter $a$. The harmonic interatomic couplings are $k$, $k'$ and $k''$ and
cubic coupling $g$ as described in the text. Arrows indicate the Fe B$_{1g}$ mode.}
\end{figure}

The temperature dependence of the lifetime and energy  of a phonon
is ordinarily due to phonon-phonon interactions that arise from anharmonic interatomic
potentials. In general the cubic anharmonicity is the dominant
term\cite{Calandra} and we consider only this.
The Raman ($\vq=0$) intensity for Stokes scattering for the mode $j$ at frequency $\omega$ is given by
$I_S(j,\omega)\propto -(1+n(\omega)) Im {\cal D}_{ret}(\vq =0,j,\omega)$ \cite{Hayes} where ${\cal D}_{ret}$ is the retarded phonon Greens function
which to linear order in the self energy $\Pi=\Delta-i\Gamma$ (minus sign by convention) gives
\begin{equation}
I_S(j,\omega)\propto \frac{(1+n(\omega))\Gamma(\vec{0},j,\omega)}{[\omega_0(\vec{0},j)+\Delta(\vec{0},j,\omega)-\omega]^2+\Gamma^2(\vec{0},j,\omega)}
\end{equation}
where $n(\omega)=(e^{\hbar\omega/k_BT}-1)^{-1}$ is the Bose occupation factor. The measured width (FWHM) is thus ideally given by $\Gamma$ and the shift by $\Delta$.
We calculate the self energy to second order in the interaction
$H_A=\frac{1}{6}\sum_{\{\vq_i\},\{j_i\}}V(\vq_1,j_1;\vq_2,j_2;\vq_3,j_3)A_{\vq_1,j_1}A_{\vq_2,j_2}A_{\vq_3,j_3}\delta_{\sum{\vq_i},0}$
where $A_{\vq,j}=a_{\vq,j}+a^{\dagger}_{-\vq,j}$ is the phonon operator.\cite{Calandra} $V$ are the matrix elements
\begin{widetext}
\begin{equation}
V(\vq_1,j_1;\vq_2,j_2;\vq_3,j_3)=\sum_{r_1,r_2,r_3,\vec{R}_1,\vec{R}_2,\vec{R}_3}\left[\prod_{i=1}^3
\left((\frac{\hbar}{2M_{r_i}\omega(\vq_i,j_i)})^{\half}e^{i\vq_i\cdot\vec{R}_i}\vec{e}_{\vq_i,j_i}(r_i)\cdot\nabla\right)\right]
\phi(r_1,r_2,r_3,\vec{R}_1,\vec{R}_2,\vec{R}_3)
\label{Vfull}
\end{equation}
\end{widetext}
where the derivatives act on $\phi$, the interatomic potential, with $r_i$ the atomic positions within unit cell $\vec{R}_i$ and $\vec{e}_{\vq_i,j_i}(r_i)$
are the displacement vectors of the respective mode. The expression thus amounts to the third derivate with respect to phonon
displacements of the atomic positions.
%Evaluating
%$V(\vec{0},Fe-B_{1g},;\vq,j_2;-\vq,j_3)$ requires in principle the complete phonon spectrum.
The iron atoms are only
weakly coupled to the $(Ce,Nd)O$ layer and since the phonon dispersions in the c-direction are weak\cite{Boeri} we can model
the the Fe-B$_{1g}$ mode by considering only the Fe-As plane. (This also makes the procedure quasi-universal as it only depends on
the properties of this plane which is common to all iron-pnictide superconductors.) For the harmonic problem we use a minimal spring model (Fig. \ref{crystal})
with nearest neighbor Fe-As potential $\frac{k}{2}\delta r^2$,
nearest neighbor Fe-Fe potential $\frac{k'}{2}\delta r'^2$, and in-plane As-As coupling $\frac{k''}{2}\delta r''^2$ with $\delta r$, $\delta r'$, and $\delta r''$
the deviations from the
respective equilibrium distances. For the Fe-As
coupling we also add a cubic term $-\frac{g}{6}\delta r^3$, where the magnitude of $g$ is to be estimated from the experimental fit.
The Raman active B$_{1g}$ is here a pure Fe mode as depicted in Fig. \ref{crystal} with energy
$\omega_{B_{1g}}=\sqrt{\frac{4k\sin^2\theta}{m}}$ ($m=56u$) and to get $\omega_{B_{1g}}=220cm^{-1}$ we take $k=8.7 eV/\AA^2$, using $\theta=35^0$.\cite{Cruz}
Values $k'=0.3k$ and $k''=0.2k$ gives a lower edge of the accoustic branches %$76cm^{-1}$ and $105cm^{-1}$ at the $X$ and $M$ points respectively
in good agrement with LDA calculations.\cite{Boeri}
Using this phonon spectrum we calculate the interaction matrix elements through Eq. \ref{Vfull} and the cubic potential.
Sampling the coupling $V(\vec{0},Fe_{B1g};\vq,j_1;-\vq,j_2)\equiv V_0(\vq,j_1,j_2)$
over the Brillouin zone shows that except at high symmetry points all 12 modes are coupled quite isotropically and does not vary
strongly with $\vq$ if the frequency dependence is included, giving
\begin{equation}
V_0(\vq,j_1,j_2)\approx\kappa g (\frac{\hbar}{2m\omega_{B_{1g}}})^{3/2}\equiv V_0\,,
\end{equation}
with the numerical value $\kappa=0.10$.
\subsection{Phonon self energy}
Now we are ready to address the frequency shift by calculating the self-energy to second order in $V_0$.
The self
energy is given by the bubble diagram\cite{Calandra}  with imaginary part
\begin{widetext}
\begin{eqnarray}
\Gamma(\omega)=\frac{\pi}{2N\hbar^2}\sum_{\vq,j_1,j_2}&&|V_0(\vq,j_1,j_2)|^2[(1+n(\omega_{\vq,j_1})+n(\omega_{\vq,j_2}))
[\delta(\omega-\omega_{\vq,j_1}-\omega_{\vq,j_2})-\delta(\omega+\omega_{\vq,j_1}+\omega_{\vq,j_2})]\nonumber\\
&+&(n(\omega_{\vq,j_1})-n(\omega_{\vq,j_2}))[\delta(\omega+\omega_{\vq,j_1}-\omega_{\vq,j_2})-\delta(\omega-\omega_{\vq,j_1}+\omega_{\vq,j_2})]]
\end{eqnarray}
\end{widetext}
(The phonon-difference process, the second term, vanishes at zero temperature but can give a significant contribution at finite temperature.)
The simplest standard way to evaluate this expression is to assume
that the scattering is diagonal in the modes $V_0\sim\delta_{j_1,j_2}$, the Klemens model,\cite{Klemens} which gives a characteristic temperature dependence
$\Gamma(\omega)\sim (1+2n(\omega/2))$. As discussed previously we find instead that the scattering
is approximately isotropic between modes and we need to do a more careful analysis.
In principle we could calculate this expression numerically at each
temperature using
our numerical phonon spectrum but instead
we will use an approximate Lorentzian fit for the phonon DOS with the advantage of giving an analytic
expression for both $\Gamma$ and $\Delta$ including the full temperature dependence.
\begin{figure}
\includegraphics[width=8cm]{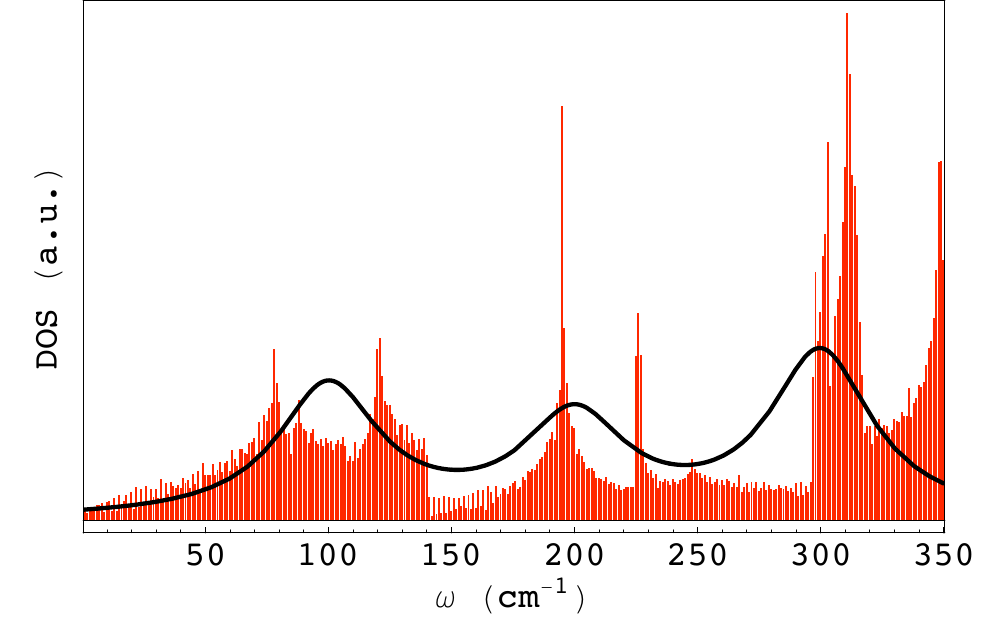}
\caption{\label{dos} (Color online) Phonon density of states of the isolated Fe-As plane calculated within the harmonic model discussed in the text and
its approximation in terms of three Lorentzians (solid curve) that contain
4,3,5 modes with increasing energy respectively. }
\end{figure}
The calculated phonon DOS, plotted in Fig. \ref{dos}, consists of three regions of high density centered around $w_1\approx 100cm^{-1}$,
$w_2\approx 200cm^{-1}$, $w_3\approx 300cm^{-1}$. The lower part contains $m_1=4$ modes, the intermediate part contains $m_2=3$
modes and the upper part $m_3=5$ modes giving
\begin{equation}
\rho(\omega)\approx \frac{N}{\pi}\sum_{i=1}^3m_i\frac{\gamma}{(w-w_i)^2+\gamma^2}\equiv\sum_i\rho_i(\omega)\,,
\end{equation}
which is thus normalized by $\int_{-\infty}^{\infty}\rho(\omega)=12N$ and where we estimate $\gamma=25cm^{-1}$.
%(We can integrate from $-\infty$ because the DOS is confined by the Lorentzians and for simplicity we use the same width $\gamma$ for all three peaks.)
To proceed we evaluate the occupation factor $n(\omega)$ at the peak position of the respective DOS and assume a q-independent
spectral weight giving $\Gamma$ as an integral
over all pairs of peaks of the spectral weight with the result
\begin{widetext}
\begin{eqnarray}
\Gamma_0(\omega)=\frac{1}{2\hbar^2}V_0^2\sum_{i,j=1}^3m_im_j&&[(1+n(\omega_i)+n(\omega_j))
[\frac{2\gamma}{(\omega-\omega_i-\omega_j)^2+4\gamma^2}-\frac{2\gamma}{(\omega+\omega_i+\omega_j)^2+4\gamma^2}]\nonumber\\
&&+(n(\omega_i)-n(\omega_j))[\frac{2\gamma}{(\omega+\omega_i-\omega_j)^2+4\gamma^2}-\frac{2\gamma}{(\omega-\omega_i+\omega_j)^2+4\gamma^2}]]\,.
\end{eqnarray}
\end{widetext}
The expression is naturally understood as the scattering of the mode at $\omega$ into two modes within the same or different DOS peaks
(or the corresponding difference process).
Note that $\Gamma_0(\omega_{B_{1g}})$ will be dominated by scattering into the low-energy peak
($i=j=1$) which will give a temperature dependence close to the Klemens model $\Gamma(\omega)\sim (1+2n(\omega/2))$.\cite{Klemens}
Using the Kramers-Kronig relation
$\Delta(\omega)=-\frac{1}{\pi}\int_{-\infty}^{\infty}\frac{\Gamma(\omega')}{\omega'-\omega}d\omega'$,
gives
\begin{widetext}
\begin{eqnarray}
\Delta_0(\omega)=\frac{1}{2\hbar^2}V_0^2\sum_{i,j=1}^3m_im_j&&[(1+n(\omega_i)+n(\omega_j))
[\frac{(\omega-\omega_i-\omega_j)}{(\omega-\omega_i-\omega_j)^2+4\gamma^2}-\frac{(\omega+\omega_i+\omega_j)}{(\omega+\omega_i+\omega_j)^2+4\gamma^2}]\nonumber\\
&&+(n(\omega_i)-n(\omega_j))
[\frac{(\omega+\omega_i-\omega_j)}{(\omega+\omega_i-\omega_j)^2+4\gamma^2}-\frac{(\omega-\omega_i+\omega_j)}{(\omega-\omega_i+\omega_j)^2+4\gamma^2}]]\,.
\label{shift}
\end{eqnarray}
\end{widetext}

\section{Discussion}
\begin{figure}
\includegraphics[width=8cm]{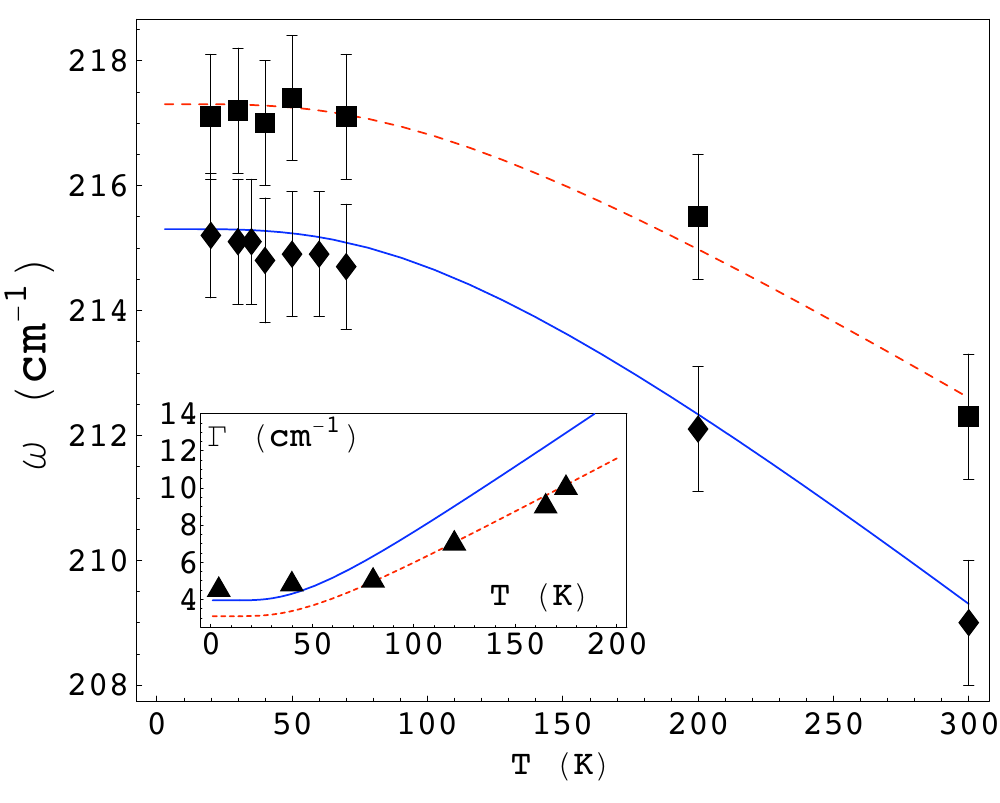}
\caption{\label{b1gfit} (Color online) Fit of the model frequency shift ($\Delta_0$) to data for $NdFeAsO_{.88}F_{.12}$ (boxes) and $CeFeAsO_{.84}F_{.16}$ (diamonds).
Inset are the corresponding model lifetimes ($\Gamma_0$) together with linewidth data (triangles) on CaFe$_2$As$_2$ from Choi et al.\cite{Choi}.}
\end{figure}

Figure \ref{b1gfit} shows the fit from equation \ref{shift} to the experimental values using $\omega=\omega_0+\Delta_0(\omega_{B_{1g}})$,
with fitting parameters $\omega_0$ and $g$ ($\Delta_0\sim g^2$). For the Nd sample we find
$\omega_0=221.7cm^{-1}$ and $g=103eV/\AA^3$ and for the Ce sample $\omega_0=221.0cm^{-1}$ and $g=116eV/\AA^3$, where $\omega_0$ is close
to the assumed harmonic value $\omega_{B_{1g}}=220cm^{-1}$.\cite{note} It is interesting to note that we find
$\Delta_0(T=0)\approx 6cm^{-1}$, i.e. even at zero temperature phonon-phonon interactions give a finite (here 3\%) contribution to the
phonon energy, an effect which LDA calculations of phonon energies based on linear response neglects.
%Frozen phonon calculations\cite{Yildirim} find an insignificant
%anharmonicity of the out of-plane modes around 200$cm^{-1}$ which is not consistent with the t and consequently not consistent with the teperature
%dependence of the energy...

Due to the polycrystalline nature of our samples we are not able to find reliable phonon widths and compare instead our theoretical results
to Raman studies by Choi {\em et al.}\cite{Choi} on CaFe$_2$As$_2$. In that material the B$_{1g}$ mode has lower energy but the temperature
dependence is similar and the width correspond well with our calculations without any additional fit.
The temperature dependence is in fact just the Klemens model but the strength of the calculation is that only one parameter $g$ gives both the shift and the width.
In Ref\cite{Choi} the linewidth variations
were tentatively assigned to changes in the electronic scattering below T$_{SDW}$. These speculations we can quite definitely rule out;
the anharmonic contribution accounts to good accuracy for both the shift and
linewidth variations of this mode.\cite{note2}

\subsection{Lattice expansion}

The cubic coupling will give rise to an expansion of the lattice that
we estimate by considering an isolated Fe-As bond for which $<0|\delta r|0>=\frac{\hbar g}{4(m*)^{1/2}k^{3/2}}$ with $m^*=Mm/(m+M)$.
We include also a nearest neighbor Fe-Fe anharmonic coupling $g'$ with the same relative strength ($g'=g(k'/k)^{3/2}$).
With this the isotope substitution $^{56}$Fe$\rightarrow$ $^{54}$Fe gives an expansion of the in-plane lattice parameter
$\Delta a=5.1\cdot 10^{-4}{\AA}$ and Fe-As distance $\Delta d_{Fe-As}=2.5\cdot 10^{-4}{\AA}$, which compared to $a\approx 4\AA$ and
$d_{Fe-As}\approx 2.5\AA$.\cite{Cruz} gives a relative expansion
$\lesssim 2\cdot 10^{-4}$.

What is the possible effect of a small lattice expansion on $T_c$ and $T_{SDW}$? There is a direct effect on the electronic hopping integrals $t$, and theories
of an isotope effect based on this has been suggested for the cuprates and C$_{60}$ as well the pnictides.\cite{Fisher,Chakravarty,Phillips}
%(Here $t$ is
%a representative hopping, with all Fe-As and Fe-Fe hoppings between different orbitals expected to get shifts of a similar magnitude.)
Assuming
$t=t_0e^{-q(r/r_0-1)}$ with $q\sim 1$\cite{comment_Vildosola} and where $\vec{r}=r_0\hat{r}+\vec{\delta}$ with small displacement $\vec{\delta}$ we find
$\frac{\delta t}{t_0}=-q\frac{\delta_r}{r_0}-\frac{q}{2}\frac{\delta_\perp^2}{r_0^2}+\frac{q^2}{2}\frac{\delta_r^2}{r_0^2}$ where $\delta_r=\vec{\delta}\cdot\hat{r}$
and $\delta_\perp=\vec{\delta}\times\hat{r}$. The linear term has only an anharmonic contribution whereas the quadratic terms will get contributions from zero point
fluctuations in the harmonic approximation. We estimate $\delta^2$ from the ground state energy per atom $\epsilon_0=38.9meV$ and $\epsilon_0=39.2meV$
for $^{56}$Fe and $^{54}$Fe respectively and energy $\epsilon_0/2$ per Fe-As bond.
The fluctuation along a bond is given by $\frac{k}{2}{\delta r}^2\approx \epsilon_0/4$ ($k=8.7eV/\AA^2$), giving the difference
$\delta_r^2\approx \Delta\epsilon_0/(2k)\approx 2.3\cdot 10^{-5}\AA^2$ and for the transverse fluctuations
$\delta_\perp^2\approx 2\delta_r^2\approx 4.6\cdot 10^{-5}\AA^2$.\cite{note_fluc}
The contribution to the shift of the hopping integrals is partially canceled by the different signs and in total smaller by an order of
magnitude compared to the anharmonic contribution.

\subsection{Isotope effect}
Consider a weak coupling (non phonon mediated) SC or SDW transition with $T_c\sim \Omega e^{-1/N(0)V}$,
where $\Omega$ is the relevant energy cut-off, which may be the magnon energy for spin fluctuation mediated pairing,
$V$ is the effective interaction and $N(0)$ the relevant electronic DOS at the Fermi energy.
Focusing on the contribution from the DOS,  $N(0)\sim 1/t$ gives the isotope exponent
$\alpha=-\frac{d\ln T_c}{d\ln m}\approx-\frac{1}{N(0)V}\frac{d\ln N(0)}{d\ln m}\approx \frac{0.5\cdot 10^{-2}}{N(0)V}\approx 10^{-2}$.
(Assuming, $N(0)V\approx 0.5$. It cannot be much smaller to get a high T$_c$.)
The interaction strength and $\Omega$ may also depend on the hopping integrals and give isotope shifts of similar magnitude.
Although this is a simple analysis
we expect the order of magnitude estimate to be relevant to any purely electronic microscopic model containing
inter and intra orbital hopping integrals and interactions that only depend indirectly on the lattice parameters.

Alternatively, we may relate the change of lattice parameter to a corresponding pressure of
$dP=\frac{3\Delta a}{a}/\beta\approx 0.4$kbar through the
compressibility $\beta=-\frac{d\ln V}{dP}\approx 1.0\cdot 10^{-3}/$kbar\cite{Zhao_JACS}
(We have no estimate of the c-axis change related to
the in-plane expansion but only assume that this is of similar relative magnitude.)
Pressure dependence of T$_c$ of around $0.2K/$kbar has been reported in several materials at different dopings although
close to optimal doping it appears that the effect is generally significantly smaller.\cite{Chu_review} Nevertheless, from these considerations
we find an upper estimate of the isotope exponent $\alpha\approx 0.06$.

Clearly something more sophisticated is needed to produce $\alpha=0.4$
as found experimentally in Ref.\onlinecite{Liu}. Encouraging perhaps, the sign from the naive weak coupling analysis based on a change in the electronic
DOS does agree with
experiments and would naturally imply a similar exponent for both SC and SDW.\cite{note3}

\section{Summary}
In summary, we find that the temperature dependent shift and width of a Raman active phonon is well represented by the anharmonic contribution,
consistent with weak electron-phonon interactions.
At the same time, we estimate the change of electronic hopping integrals due to Fe isotope substitution and find that these are too small to
generate an
isotope effect on T$_c$ or T$_{SDW}$ of the magnitude reported in \cite{Liu} without an explicit phonon contribution.
These apparently contradictory results present a
significant challenge for any theory of superconductivity in the iron-pnictides.

\section{Acknowledgement}
We thank Professor Nan Lin Wang for providing the samples used in the Raman measurements.\\

{\em Note added} After the submission of this work x-ray diffraction (XRD) data on the samples used for the isotope experiments appeared
(Ref. \onlinecite{Liu}, supplemental).
Changes in the lattice parameter of isotope substituted samples are the same within the experimental error of $\sim 1\cdot 10^{-3}\AA$, which is
greater than the change $\Delta a\approx 5\cdot 10^{-4}\AA$ calculated here.
There also appeared a report of a negative Fe-isotope effect with $\alpha=-0.18$.\cite{negative_iso} Again, XRD data find the same in-plane
lattice parameter within experimental error of $\sim 1\cdot 10^{-3}\AA$.\cite{Shirage_private}


\begin{thebibliography}{12}
\bibitem{note_undoped} We performed measurements also on undoped ($x=0$) crystals of the same materials. Apart from small changes
in phonon energies we found no significant difference in the Raman spectra between doped and undoped samples.
\bibitem{note} Note that $\Delta_0(\omega)$ is
a quite rapidly varying function of $\omega$ and sensitive to the input parameters $\omega_i$ and $\gamma$.
Even avoiding any fine tuning of parameters we find that $g$ can vary by up to 50\% from the values given above for moderate changes of our model DOS.
However, values of $g\sim 100 ev/\AA^3$ are also in line with estimates for GaAs\cite{Pietsch} which has a similar local
lattice structure.
\bibitem{note2} In \cite{Choi} emphasis was put on changes at the SDW transition. While our results match that data for the low temperature
orthorombic phase we have no explanation for the behavior above T$_{SDW}$ where the B$_{1g}$ linewidth saturates and the As A$_{1g}$ mode has
unexpectedly vanished.
\bibitem{comment_Vildosola} According to LDA calculations with varying lattice parameters\cite{Vildosola} the relative changes of the bandwidth are approximately the
same as the relative changes of the lattice parameters.
\bibitem{note3} The simplest strong coupling
analysis of the SDW order on the other hand in terms of a large on-site interaction $U>>t$ would have
$T_{SDW}\sim J=t^2/U$, again a tiny but in this case negative isotope effect.
\end{thebibliography}
\end{document}